\documentclass[preprint,12pt]{elsarticle}
\usepackage{graphicx, nicefrac}
\usepackage{siunitx}
\usepackage{bm}
\usepackage[version=4]{mhchem}
\usepackage{chemformula}
\usepackage{color,hyperref}
\usepackage{float}
\usepackage{soul}
\usepackage[switch]{lineno} 
\begin{document}
\begin{frontmatter}
\begin{abstract}
    Progress in information processing relies on spintronics, where magnetic states serve as efficient carriers for data storage and transfer. In this work, we theoretically study magnon propagation in a bilayer composed of a ferromagnet and an antiferromagnet. For this purpose, we probe the spin Seebeck effect by introducing a spatially varying temperature profile. This generates a local magnon excitation and a continuous magnon flux from hot to cold regions which we quantify through the resulting non-equilibrium magnon accumulation. Based on the chirality of these modes, we identify specific constraints for magnon modes traveling either from the ferromagnet into the antiferromagnet or vice versa. A key finding is the observation of a thermally triggered spin current in the antiferromagnet, a phenomenon typically absent in bulk antiferromagnets that obey time-reversal symmetry. These results provide important insights into the design of heterostructures for magnonic chirality-selective spin transport.
\end{abstract}

\title{Thermal magnon transport in FM/AFM bilayers}
\author{Moumita Kundu}
\affiliation{organization={Fachbereich Physik, Universität Konstanz},
addressline= {Universitätsstraße 10},
city = {Konstanz},
postcode = {78464},
country = {Germany}}
\author{Ulrich Nowak}
\affiliation{organization={Fachbereich Physik, Universität Konstanz},
addressline= {Universitätsstraße 10},
city = {Konstanz},
postcode = {78464},
country = {Germany}}

\date{\today{}} 


\end{frontmatter}

\section{Introduction}
With ever-increasing demand for information processing, integrating magnetic functionalities into  electronic devices is a promising approach \cite{Chumak2015,magnonic_roadmap_ABarman}. In this context, the investigation of excitations of an ordered magnetic state is crucial. Such excitations are referred to as spin waves and in the quantized form known as magnons \cite{Kittel_magnons}.  The behavior of magnons in a specific magnetic environment is of immense importance for the optimization of future spintronics devices \cite{VVKruglyak2010}. When used as carriers of information instead of electrons, magnons enable the use of insulating magnetic materials rather than conductors, thereby reducing Ohmic heating and electrical power consumption  \cite{J_Cramer_spin_valve,BRATAAS20201,sw_logoic_gates,magnon_transistor}. 
    
The spin-Seebeck effect describes the fact that a thermal gradient can give rise to spin currents \cite{Adachi_2013}. These spin currents can be detected electrically via the inverse spin-Hall effect \cite{Saitoh2006}, using non-magnetic Pt stripes that convert the spin current into an electrical charge current \cite{Dyakonov2008}. Meanwhile, the spin Seebeck effect has been observed in a variety of bulk materials, including magnetic conductors \cite{Uchida2008}, semi-conductors \cite{Jaworski2010}, insulators \cite{SSE_insulator} and even half-metallic Heusler compounds \cite{SSE_magnon_phonon_mediated}.

It is well established that an  antiferromagnet (AFM) usually exhibits an imperfect degree of sublattice magnetization compensation at the interface with an adjacent ferromagnet (FM), an effect that can be used for the transformation of spin currents \cite{PhysRevB.93.224421}. Exploiting this effect, a YIG/CoO/Co heterostructure was designed recently, acting as a spin-valve structure, where magnonic spin current can be switched via the orientation of the two ferromagnetic layers \cite{J_Cramer_spin_valve}. It was found that the detected spin current emitted by resonant spin pumping depends on the relative alignment --- parallel or antiparallel --- of the YIG and Co magnetization. Local excitation of specific magnon modes in a GdIG and NiO bilayer and trilayers revealed that transmitted mode properties, such as penetration length and amplitude, depends on the excited frequency and polarization along with the interface coupling \cite{Wang2025,Wang2014}. Proximity induced magnons were also investigated in FM/AFM/FM trilayers where even in thermal equilibrium the adjacent FMs could induce antiferromagnetic order in the central layer \cite{Brehm2022Magno-57193}. 
Collectively, these studies indicate that FM/AFM interfaces can act as polarization dependent magnonic spin filters, inducing properties beyond those of the individual magnetic films. 

In this work, we investigate the thermal transport of magnons in bilayers of an FM in contact with an AFM, focusing on the filtering of magnon modes due to the distinct properties of the spin wave dispersions in the two materials and nonreciprocal transport.
We use two setups, one where the FM is thermally excited and the magnon propagation is measured in the AFM and one vice versa, where the AFM is excited and transport in the FM is investigated. 
Normally, when the two magnonic bands of the AFM are not split, one would not expect any thermal transport of angular momentum via magnons in the bulk of the AFM since the contributions of the two bands cancel each other \cite{Ohnuma2013}. In our setup, however, due to the unidirectional flow of chiral magnons from the FM into the AFM, a spin accumulation is observed in the AFM and a finite spin current is found in the absence of any relativistic effects or applied magnetic field \cite{PhysRevLett.115.266601,PhysRevLett.116.097204,PhysRevLett.122.217204}. This net magnetization can be linked to the lifted degeneracy of the AFM modes, due to the nonreciprocal magnon transport originating from proximity effects.

\section{Method: Atomistic spin dynamics} \label{sec:method}
In the following, we use a semi-classical approach where the spin operators of a  Heisenberg Hamiltonian are replaced by normalized spin vectors $\mathbf{S}_i$,
    \begin{equation}\label{eqn:hamiltonian}
        \mathcal{H} = - \sum_{i \neq j}^N \frac{J_{i,j}}{2} \mathbf{S}_i \cdot \mathbf{S}_j - d_{z} \sum_{i=1}^N (S_i^z)^2 .
    \end{equation}
Here, the first term represents an isotropic exchange interaction with nearest neighbor (NN) exchange constant $J_{i,j}$ and the second term represents a uniaxial anisotropy with anisotropy constant $d_{z}$, which prefers a ground state spin configuration aligned with the $\pm \hat{z}-$axis. In our simulations, we use $J_{\text{FM}}$=\SI{10}{\milli\electronvolt} (resembling Cr$_2$Te$_3$ and Fe$_3$GeTe$_2$), $|J_{\text{AFM}}| =$ \SI{20}{\milli\electronvolt} (similar to MnTe, Cr$_2$O$_3$ and FeSn), and $d_z =$ \SI{0.01}{\milli\electronvolt}. The exchange interaction induces an ordering temperature of $\approx$ \SI{170}{\kelvin} for the FM and $\approx$ \SI{340}{\kelvin} for the AFM. 

Spin dynamics is determined via the stochastic Landau-Lifshitz-Gilbert (s-LLG) equation \cite{LL1935,Nowak2007classical,Gilbert_eqtn,Brown1963,Brown1979},
    \begin{equation}\label{eqn:llg}
        \frac{d\mathbf{S}_i}{dt} = -\frac{\gamma}{\mu_s(1+\alpha^2)}\big(\mathbf{S}_i \times \mathbf{H}_{\textrm{eff},i} + \alpha\mathbf{S}_i\times(\mathbf{S}_i\times\mathbf{H}_{\textrm{eff},i}) \big).
    \end{equation}
Here, $\gamma=1.79\times 10^{11}$ Hz/T is the gyromagnetic ratio and $\mu_s=2\ \mu_B$ is the atomic magnetic moment, with Bohr's magneton $\mu_B$. The first cross product describes the precession of spins around their  effective field and the second term, with the double cross product, describes their relaxation towards the direction of the effective field.  The timescale of this relaxation depends on the strength of the Gilbert damping constant, $\alpha$, which is chosen to be 0.008 to keep the magnon propagation lengths within the finite size of the simulated volume.

For each spin the effective field, $\mathbf{H}_{\textrm{eff},i}$, is defined  as
    \begin{equation}\label{eqn:effective-field}
        \mathbf{H}_{\textrm{eff},i} = -\frac{\partial \mathcal{H}}{\partial \mathbf{S}_i} + \boldsymbol{\xi}_i .
    \end{equation}
To model thermal properties, $\boldsymbol{\xi}_i$ is a noise term in the form of a Gaussian white noise that fulfills the fluctuation-dissipation theorem and is uncorrelated in time and space with \cite{Brown1979}, 
    \begin{subequations}
	       \begin{align}
	 	         \langle \xi_i(t) \rangle &= 0 \\
	 	         \langle \xi_{i\eta}(t)\xi_{j\theta}(t^{\prime}) \rangle &= 2\alpha k_BT_i\mu_s\delta_{i,j}\delta_{\eta,\theta}\delta(t-t^{\prime})/\gamma .
	       \end{align}
	\end{subequations}
Since we are interested in thermal gradients, the temperature $T_i$ of the heat bath is coupled to the spin system, and can vary in space in terms of a temperature profile. Indices $\eta,\theta$ represent the spatial coordinates, whereas $i, j$ denote lattice sites. Note, that the s-LLG equation of motion is non-linear in the spin degrees of freedom and, hence, captures magnon-magnon interactions. The simulations assume single heat bath temperature where the electronic and lattice degrees of freedom are not distinguished. The spin system is free to equilibrate with its heat bath, though no assumption is made regarding the existence of a spin temperature.
    \begin{figure}[t]
        \centering
    \includegraphics[width=0.75\linewidth]{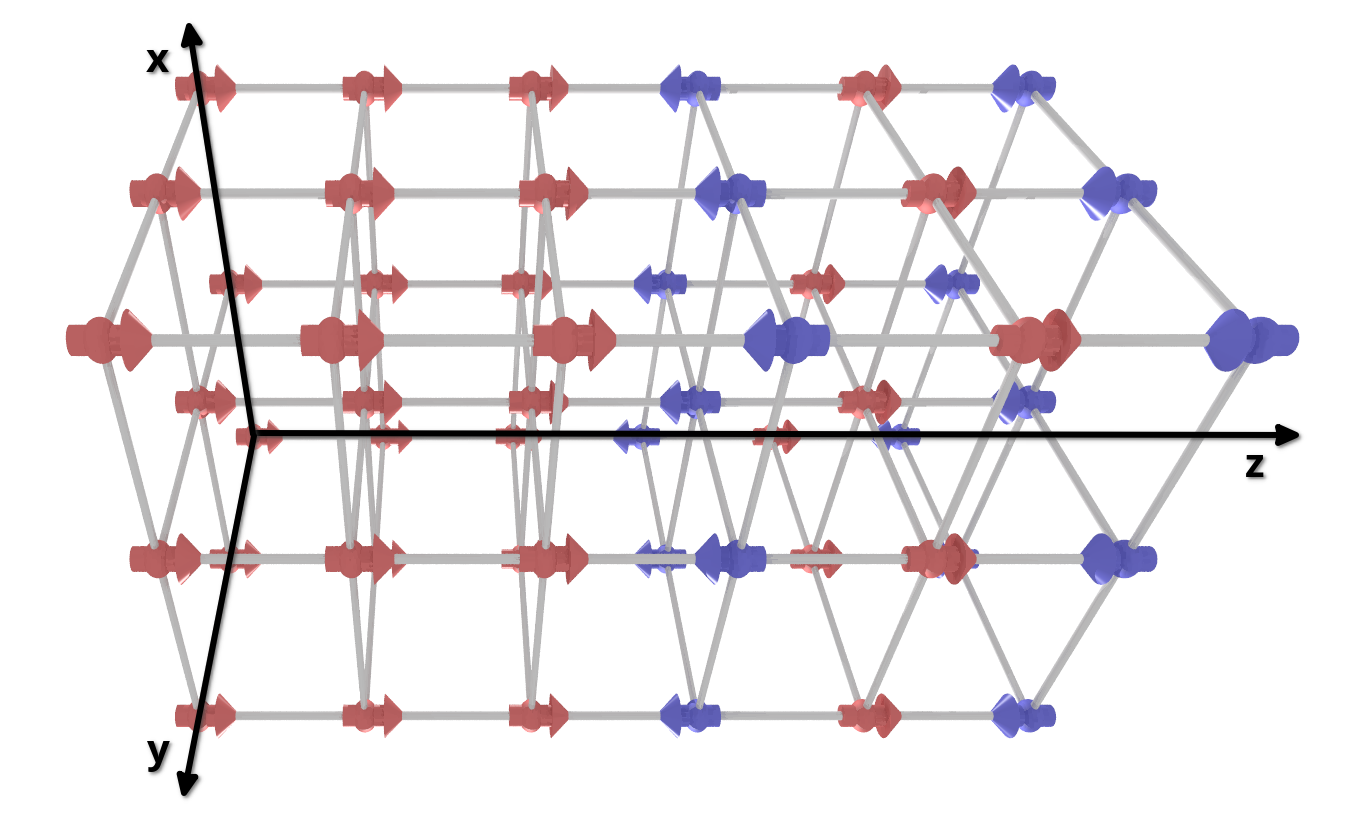}\quad
        \caption{Sketch of the magnetic bilayer as implemented in our simulations, where the left part with all spins pointing to the right ($+\hat{z}$) models the FM and the right part models a layered AFM with spin pointing alternately left and right.}
        \label{fig:bilayer_structure}
    \end{figure}

A sketch of the simulated model is presented in Fig. \ref{fig:bilayer_structure}, where the left layers represent an FM with all spins pointing along $+\hat{z}$ direction and the layers on the right represent a layered AFM with alternating spin structure (l-AFM). Here the interface is fully uncompensated with negative interface coupling $J_{\text{AFM}}$. The magnetic unit cell of the FM has a lattice constant of $a$ and that of the AFM is $2a$ as it contains two subsequent layers, A and B, with alternating spins. The number of layers in $z$-direction will be denoted as $r_z$ for further reference. Simulations are performed for a system size of $8 \times 8 \times 256$ spins where layers along $\hat{z}$ are divided equally among the FM and the AFM. The sample has periodic boundary conditions along $\hat{x}$,  $\hat{y}$ directions and open along $\hat{z}$.
When spin transport along $\hat{z}$ is investigated, reflections of magnons from the sample's open boundary have to be avoided. To achieve this, absorbing boundary conditions are implemented at the colder end of the sample, where the damping constant in the last layers increases exponentially to damp all excitation.

For any finite temperature $T$, magnons appear naturally as thermal excitation in our spin model. With the ground state spin configuration along $\hat{z}$ direction, we compute the magnon dispersion from the complex spin-wave components $S_{\text{sw}}(t,r_{z})=\langle S_{x}(t,r_{z})\rangle + i \langle S_{y}(t,r_{z})\rangle$.
Here $r_{z}$ represents the magnetic unit cell (as described above) along $\hat{z}-$direction and an average over the perpendicular $(xy)-$plane is taken, indicated by the angular brackets. 
Along $\hat{z}$-direction, $S_{x}$ and $S_{y}$ represent each magnetic unit cell. For the FM, this is a single monolayer, while for the AFM, it is the sum of two sublattices with antiparallel ground state spin configuration.
We perform a two-dimensional Fourier transform in time (from 0 to our simulation time $t_\mathrm{max}$) and space (from 0 to our system size along $z$-direction) to get the spin wave intensity,
\begin{equation}
    I\left( f,k_z \right) =\bigg \lvert \frac{1}{2\pi}\int_0^{t_\mathrm{max}} \int_0^{L_z} S_{\textrm{sw}}\left(t,r_z \right) e^{-i\left(2\pi f t -   k_{z}r_{z} \right)} dt\,dr_{z} \bigg \rvert ^2,
\end{equation}
which also reflects the magnon dispersion along $\hat{z}$ direction. This Fourier transform is averaged over an ensemble of 10 simulations to improve the analysis for both, equilibrium and non-equilibrium conditions. It is performed individually for the FM and AFM. Additionally, we determine the polarization of the excited modes from the sign of the $f$ and we will present left and right handed modes in the following accordingly. This connection between polarization and sign of frequency holds, since --- in the classical approximation --- magnons can be represented as plane waves with,
\begin{equation}
\begin{aligned}
S_i^x(r_z,t) &= \sum _{k}S_0 \cos(2\pi f t - k r_z) \\
S_i^y(r_z,t) &= \sum _{k}S_0 \cos(2 \pi f t - k r_z + \phi).
\end{aligned}
\end{equation}
For circularly polarized magnons the phase shift is $\phi=\pm \pi/2$ and the second line follows a $\pm \sin(2\pi f t)$ function. Consequently, the two frequencies $\pm f$ represent counter-clockwise and clockwise precession of the spins.

To study the spin Seebeck effect, thermal gradients are applied --- either in the form of a step function or linear profiles ---  which create an inhomogeneous magnon distribution in the sample leading to a non-vanishing magnonic chemical potential. A non-equilibrium magnon accumulation is thus created, which triggers a magnon current. Since thermal magnons are continuously pumped from the hotter towards the colder regions of the sample, both, magnon accumulation and magnon current achieve a quasi-static state after a relaxation time period which depends on the damping parameter. These quantities are then analyzed taking time averages and spatial averages in  planes perpendicular to the transport directions to obtain spatial profiles. Most importantly, combining these techniques above, the connection between magnon transport and nonreciprocal magnon dispersions can be established.

\section{Results} 
    \subsection{Equilibrium properties of the bilayer}
    \label{sec:results_equilibrium}
For a better understanding of the non-equilibrium transport properties later on, we start the analysis of our FM/AFM bilayer with their equilibrium dispersion relations.
\begin{figure}[t!]
    \centering
    \hspace{-1mm}
    \includegraphics[scale=0.75]{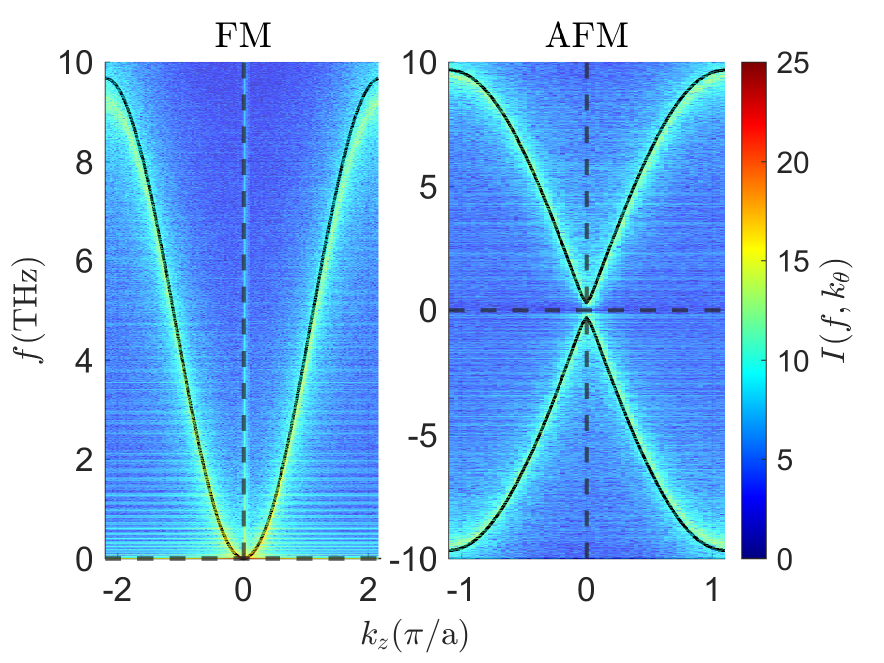}
    \caption{Magnon dispersion for thermal equilibrium conditions, on the left for the ferromagnet, with only one mode, and on the right for the antiferromagnet, with two degenerate modes possessing opposite polarization. }
    \label{fig:equilibrium_magnon_dispersion}
\end{figure}
In Fig. \ref{fig:equilibrium_magnon_dispersion} the magnon dispersion is shown for both, the FM as well as the AFM. The spin waves are circularly polarized and the FM is seen to have only one helicity, shown as positive frequencies, whereas the AFM shows both helicities, shown as positive and negative frequencies. 
Note that the Brillouin zone of the AFM is half compared to the FM since its unit cell contains two atoms. 
    
In agreement with the numerical data, the black solid lines show the analytical solutions for the magnon dispersion relations for both, FM as well as AFM. These expressions are calculated via linear spin-wave theory based on the Hamiltonian given in equation \ref{eqn:hamiltonian}, and assuming small deviations from the ground state, such that $S_i^z \approx 1$. This calculation yields the following dispersion relations for the FM,
    \begin{subequations} \label{eqn:dispersion_relations}
        \begin{equation}
        \centering
            f_{\text{FM}} = \frac{\gamma}{\mu_s(1+\alpha^2)} \Big(2d_z + 2J_{\textrm{FM}}\sum_{k_z}\big(1 - \text{cos}(k_{z}a)\big) \Big) ,
        \end{equation}
        and the AFM,
        \begin{equation}
        \begin{aligned}
            f_{\text{AFM}} &= \frac{\gamma}{\mu_s(1+\alpha^2)}\Big((2|J_{\textrm{AFM}}| + 2d_z)^2 \\
            & - (1+\alpha^2)\big(2|J_{\textrm{AFM}}|\sum_{k_z} \text{cos}(k_z a)\big)^2 \Big)^{1/2}.    
        \end{aligned}
        \end{equation}
    \end{subequations}
From Eqs. \ref{eqn:dispersion_relations}, we see that the band gap for $k \to 0$ is  $f_{\textrm{0,FM}} \approx \frac{\gamma}{\mu_s}2d_{z}$ for the FM, assuming low damping. For the AFM it is $f_{\textrm{0,AFM}} \approx \frac{\gamma}{\mu_s}\left(4d_{z}^2 + 8J_{\textrm{AFM}}d_{z} \right)^{1/2}$ and, assuming low anisotropy $d_{\textrm{z}} \ll J_{\textrm{AFM}}$, it is $f_{\textrm{0,AFM}} \approx \frac{\gamma}{\mu_s}\sqrt{8J_{\textrm{AFM}}d_{z}}$. The band gap in Fig. \ref{fig:equilibrium_magnon_dispersion} is most prominent for the AFM, where it is significantly larger than in the FM. The maximum frequencies at the edge of the Brillouin zone can also be derived from the analytical solution above as $f_{\textrm{max,FM}} \approx \frac{\gamma}{\mu_s}(2d_{z} + 4J_{\textrm{FM}})$ and $f_{\textrm{max,AFM}} \approx \frac{\gamma}{\mu_s}\left(2|J_{\textrm{AFM}}| + 2d_{z} \right)$. For our choice of model parameters these two frequencies are identical as evident at the zone edge in Fig.\ref{fig:equilibrium_magnon_dispersion}.

In the classical limit that we discuss   here, the spin wave excitations are not discrete but can transport a continuous amount of angular momentum. Thus the probability of a mode being occupied is given by the Rayleigh-Jeans distribution, $n(f,T) = \frac{k_BT}{h f}$. 
Consequently, in thermal equilibrium the high frequency modes are excited with higher probability as compared to a Bose-Einstein distribution, an effect which will be less relevant in the following since we are mostly interested in non-equilibrium properties.  In equilibrium, however, all energetically feasible modes are excited equally for symmetric directions with identical amplitudes for $\pm k$-directions  leading to no net flow of magnons in real space. This will change for nonequilibrium situations as discussed below.

\subsection{Magnon injection from FM to AFM}
\label{sec:noneq_SSE}
To investigate the spin Seebeck effect \cite{Adachi_2013} in our bilayer, we assume a temperature gradient in the form of an abrupt temperature step that excites the FM only and leads to a propagation of magnons into the AFM. The AFM is kept at \SI{0}{\kelvin} to ensure that any signal on the colder side is due to incoming magnons injected from the FM. Although this temperature profile is unrealistic in any experimental situation, it is the best defined model to study transport properties in our bilayer.       
    \begin{figure}
    \centering
    \includegraphics[width=0.75\linewidth]{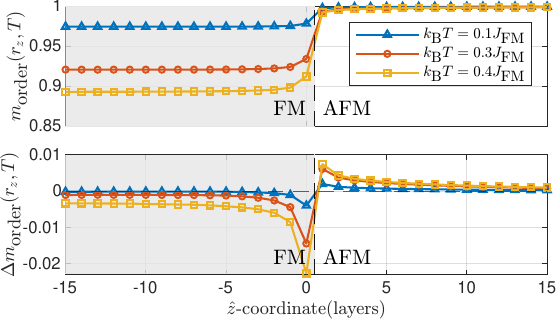}\quad
     \caption{Order parameter profile (top) and magnon accumulation profile (bottom) for different temperatures $T$, where the FM is thermally excited and the AFM is at 0K.}
     \label{fig:spatial_FA}
     \end{figure}
     
In Fig. \ref{fig:spatial_FA} (top), order parameter profiles, $m_{\textrm{order}}(r_z,T)$, are shown for different temperatures. For the FM, the latter is defined as the easy-axis magnetization component, $m_{\textrm{order,FM}}= \langle S_{z}(r_z,T)\rangle_{xy,t}$ averaged for each normal layer $\langle \rangle_{xy}$ and then time averaged $\langle \rangle_{t}$ to obtain equilibrium properties. For the AFM, the order parameter is the corresponding N\'{e}el vector $m_{\textrm{order,AFM}}=n_{z}(r_{z},T)=\langle(\langle S_{\textrm{z,A}}(r_{z},T)\rangle_{xy} - \langle S_{z,\textrm{B}}(r_{z},T)\rangle_{xy})/2\rangle _{t}$. The heated region is on the left and slightly shaded, the black dashed line indicates the interface between the FM and the AFM, where the spins couple antiferromagnetically. Due to the enhanced  local temperature on the left, the magnetization is reduced as one increases the temperature in the FM. This reduction can be interpreted as increasing number of magnons. The cold region at \SI{0}{\kelvin} on the right has higher order parameter values saturating to the ground state order parameter value. Since the two regions are coupled, the order parameter profile develops in the stationary state with an intermediate transition region which reflects finite magnon temperature on both sides of the interface. To quantify the local magnon population due to finite magnon temperature, we define the magnon accumulation as in \cite{Ulrike_2SLmagnet}, 

\begin{equation}\label{eqn:magnon_accumulation}
    \Delta m_{\text{order}}(r_z,T) = m_{z,\text{eq}}(T) - m_{\text{order}}(r_z,T).
\end{equation}
This accumulation is the difference between the actual order parameter profile $m_{\text{order}}(r_z,T)$
and the temperature dependent bulk equilibrium order parameter $m_{z,\text{eq}}(T)=\langle S_{z}(t,T)\rangle_{xyz,t}$ which is computed in a separate simulation for a homogeneous heat bath. The results are presented in Fig. \ref{fig:spatial_FA} (bottom). 
Note again, that the magnon accumulation is computed based on the respective order parameters for both the constituents.

\begin{figure}[t!]
   \centering
    \includegraphics[scale=0.9]{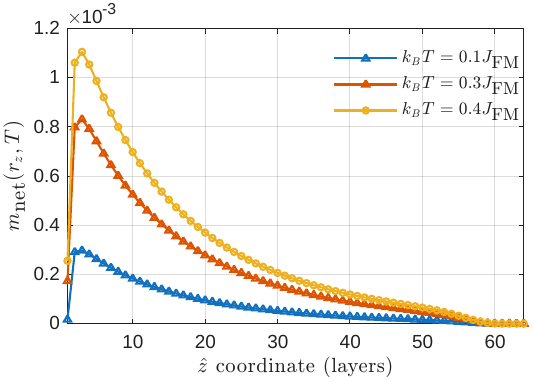}
    \caption{Net magnetization profile of the bilayer zoomed at the AFM. A finite net magnetization is found.}     
    \label{fig:spatial_total}
\end{figure}
The magnon accumulation in Fig. \ref{fig:spatial_FA} shows a negative peak in the last heated FM layer corresponding to a maximum reduction of the magnon density, since hot magnons move constantly through the interface into the colder AFM. This negative peak is also in line with a locally higher magnetization because of a loss of magnons close to the interface to the colder AFM.
Accordingly, the next layer --- the interfacial AFM layer --- shows a positive peak, corresponding to a maximum increase in magnon density compared to the colder bulk region where both, magnon accumulation and magnon density should vanish at \SI{0}{\kelvin}.  In this colder region, the magnon accumulation decays exponentially, determining the propagation length of the magnons which was previously shown for FMs and 2-sublattice ferrimagnets in \cite{Ulrike_2SLmagnet,Ritzmann2014}. 
Over all, order parameter and magnon accumulation profiles can collectively serve  as signatures of magnon transport and population.

The temperature (step) dependence of the magnon accumulation in Figure \ref{fig:spatial_FA} shows an enhanced magnon accumulation for increasing temperature, mostly in the vicinity of the temperature step. This is in line with the magnetization gradient acting as driving force for the diffusion of magnons. This is seen prominently in the FM where the magnetization is reduced for higher gradients. The accumulation in the colder region is clearly lower than in the hotter one. This asymmetry in accumulation magnitude is consistent with later results and is attributed to reflections from the interface back into the FM. This effect will be discussed in connection with the non-equilibrium magnon dispersions later on. 

The accumulation in the AFM that was shown so far was that of the N\'{e}el order parameter. Interestingly, even a finite net magnetization, $m_{\text{net}}(r_z,T)=\langle(\langle S_{z,\text{A}}(r_{z},T)\rangle_{xy} + \langle S_{z,\text{B}}(r_{z},T)\rangle_{xy})\rangle _{t}$ is obtained and displayed in Fig. \ref{fig:spatial_total}. Though the order of magnitude is significantly lower by a factor of 10 as compared to the magnon accumulation in Fig. \ref{fig:spatial_FA}, to find a net magnetization at all in a fully compensated AFM, without any relativistic effects is a crucial outcome. This net magnetization has a similar temperature dependence as the magnon accumulation shown in Fig.  \ref{fig:spatial_FA} which intuitively connects these two quantities as a result of the magnon transport triggered by the temperature gradient. 

To understand the occurrence of a net magnetization in the AFM, we note that the  AFM can hold magnons with both the polarizations, left-handed (LH)  as well as right-handed (RH), as shown before in Fig. \ref{fig:equilibrium_magnon_dispersion}. These two polarizations are associated with magnons of opposite spin and corresponding magnetization. 
Magnons in the FM, however, have only one polarization. 
Moreover, the FM couples to the AFM via the RH magnon modes only, which are permitted in the AFM but no LH magnons are excited. This leads to an imbalance of magnon mode occupations  in the AFM and the two spin directions no longer cancel each other. Consequently, a net magnetization appears. A similar effect explains the spin valve effect in \cite{J_Cramer_spin_valve}.
In the following, we will investigate this effect in depth in terms of the non-equilibrium dispersion relations.

\begin{figure}[t]
   \centering
   \includegraphics[scale=0.75]{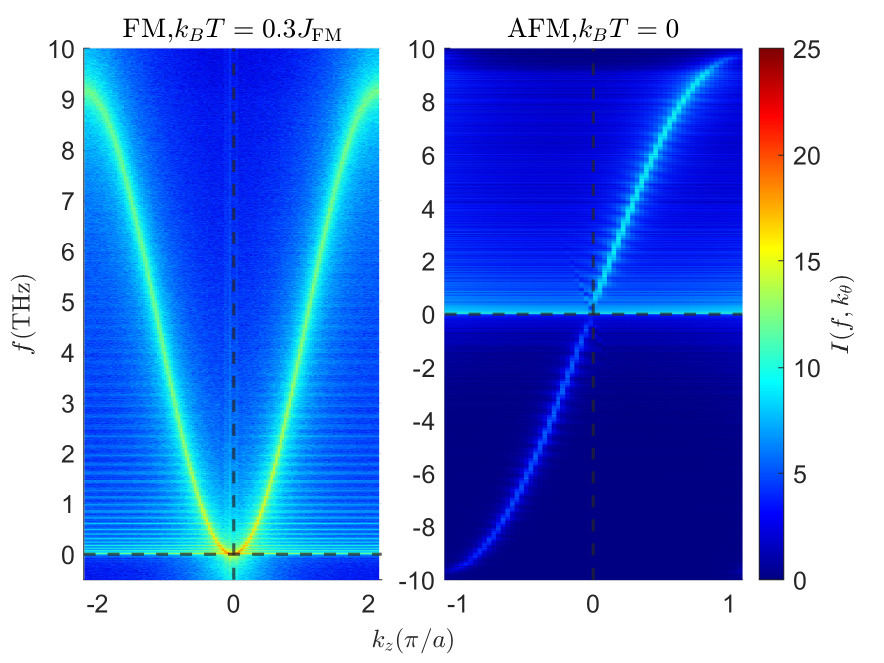}
   \caption{Non-equilibrium magnon dispersion of the bilayer where the FM (left) is thermally excited and magnons are injected into the AFM (right).}
   \label{fig:FM_AFM_2d}
\end{figure}

Fig. \ref{fig:FM_AFM_2d} shows the magnon dispersion for our bilayer calculated in the same spin Seebeck set up discussed before. On the heated ferromagnetic side, only one RH magnon mode is observed  \cite{magnon_fm_afm}. In the FM, the thermally excited modes can travel in any direction in space resulting in equal intensities along $\pm k$-directions. The AFM is at zero temperature and magnons are excited only locally by magnon diffusion from the hotter FM. This leads to significant changes of the intensities in AFM magnon dispersion. Most prominent is now only one branch, namely the one which matches the RH polarization of the FM that excites these spin waves, $+\omega_{+k}$. Additionally, it has positive group velocity, $v_{k}=\frac{\partial \omega_{k}}{\partial k}$ indicating a magnonic motion away from the FM. Branches with negative group velocity do not appear. The predominance of the group velocity selecting the mode that should propagate into the AFM shows the significance of chirality-selective behavior where the polarization and direction of travel are taken into account. The main observation here is $\omega_{+k} \neq \omega_{-k}$ leading to the RH-mode being asymmetric and chiral.

The LH mode in the AFM with positive group velocity has significantly reduced intensity compared to the RH  mode. This negative branch is most prominent in our simulations for an enhanced thermal gradient and when the temperature step is closest to the interface. When we shifted the temperature step further inside the FM (not shown in the paper), the intensity of the negative branch is reduced further. We conclude, that it is excited by local thermal fluctuations in the FM that couple to the AFM. However, the imbalance of magnons with opposite polarization leads to the net magnetization observed in Fig. \ref{fig:spatial_total}. 

The band gaps remain unchanged in both cases, but bright horizontal lines appear in the band gap of the  AFM below the lowest excited frequency of the RH magnon branch. These frequencies are not allowed in the AFM but --- because of the lower band gap of the FM --- they do occur in the FM. Due to the coupling of the FM to the AFM, they enter the AFM as evanescent modes which are immediately damped and lead to the bright lines in the Fourier spectrum. Furthermore, these modes are  partially reflected at the interface back into the FM, leading to the bright lines near the FM band gap. 
This aligns with previous findings revealing frequency dependent amplitude of transmitted modes in such a bilayer \cite{Busel2019}. 

Over all, we find a non-reciprocal, unidirectional flow of magnons from the hot FM into the colder AFM, where predominantly the RH modes of the adjacent FM occur. This makes the bilayer structure behave as a spin-wave filter, where only certain modes pass on to the AFM and the restricted modes which are not allowed in the AFM are reflected back inside of the ferromagnet. This filtration of modes leads to lifted degeneracy leading to the net magnetization induced in the AFM. Taking into account that --- in a device --- the FM could be switched, which controls the polarization of the FM, we can refer to our finding as chirality-selective magnon propagation \cite{Chirality_dependent_spin_transport,Li2025}, and have demonstrated the impacts of this selectivity as induced magnetization in a compensated AFM.

\subsection{Magnon injection from AFM to FM}
\label{sec:noneq_SSE}
\begin{figure}[!t]
\centering
\includegraphics[width=0.75\linewidth]{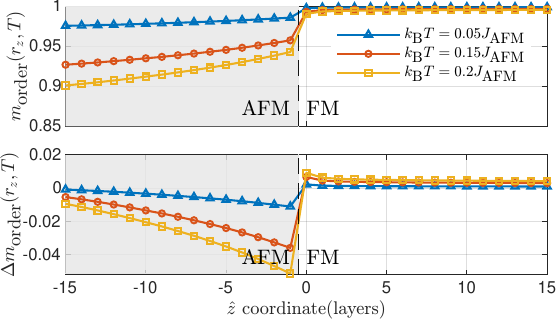}
\caption{Order parameter profile (top) and magnon accumulation (bottom) for different temperatures $T$, where the AFM is thermally excited and the FM is at 0K.}
\label{fig:bilayer_rev_macc}
\end{figure}
             
To explore the role of chirality for magnon transport further, we investigate the reversed spin Seebeck setup and excite the AFM thermally, triggering magnons propagating from the AFM into the FM. 
The excitation mechanism is the same as above, as well as the analysis. The magnetic order parameter profile along with the  magnon accumulation is plotted for this reversed geometry in Fig. \ref{fig:bilayer_rev_macc}. Analogously, the order parameter decreases in general for increasing temperatures. Near the temperature step, it shows an exponential equilibration profile defining the magnon propagation length. The magnetic order parameter increases in the AFM and reduces in the FM, demonstrating the onset of diffusion. 

The magnon accumulation shows similar properties like the previous setup in Fig. \ref{fig:spatial_FA}. The accumulation is negative in the hot region and positive in the cold one, implying on the one side a lack and on the other a surplus of magnons. An additional asymmetry between the two peak accumulations (the enhanced lack of magnons in the AFM as compared to those propagating into the FM) is attributed to the reflection of magnons at the interface. 

The magnon propagation length is visibly larger in the AFM as compared to the FM (in Fig. \ref{fig:bilayer_rev_macc}), an effect which is also known from the analytical calculations in \cite{Cramer2018,Kehlberger2015}, where it was shown that the propagation length for a FM is maximum at the band gap and reduces thereafter, while in an AFM, the propagation length first increases rapidly away from the band gap and then decreases for higher frequenices. Thus, even in a similar frequency range, the propagation lengths differ for both the materials. What is interesting to note here is the proximity driven magnetization in the AFM which will now also be seen in the magnon dispersions. 
 
\begin{figure}[t!]
\centering
\includegraphics[scale=0.75]{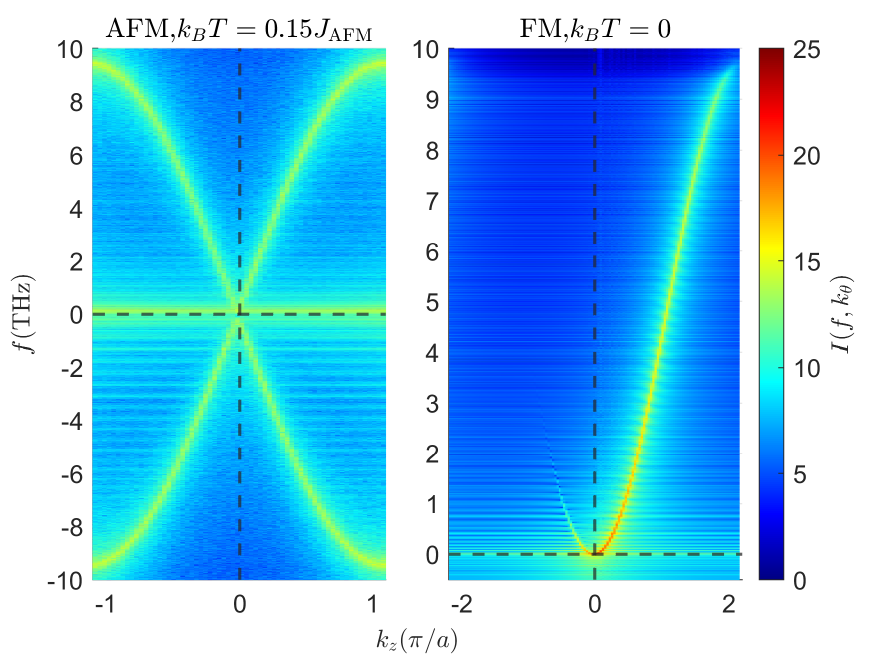}
\caption{Non-equilibrium magnon dispersion relation of the bilayer when the AFM is thermally excited and magnons are injected into the FM.}
\label{fig:2D_AFM_FM}
\end{figure}

The dispersion relation presented in Fig. \ref{fig:2D_AFM_FM} shows that first of all, in the AFM both, RH and LH magnon modes appear, while in the FM --- since only one polarity is permitted ---- the positive frequency branch along $+k$ direction exists solely. This again illustrates the unidirectional flow of magnons in the FM. Note also, that this magnon mode fades slightly at the edge of the Brillouin zone. This is attributed to the FM being only indirectly heated via the incoming magnons, which is insufficient to excite the high frequency modes. 

There is also a small tail in the lower frequency range $(\omega_{-k})$ that is excited locally from the FM modes. Since these modes are locally excited due to the finite magnon temperature in the FM they propagate in both directions. However, these low frequency magnons are within the band gap of the AFM and they appear as evanescent modes, visible as bright horizontal lines near $f=f_{0,\mathrm{FM}}$. A similar magnon blocking was observed in an FM1/AFM/FM2 trilayer, where the propagation probability was controlled via the orientation of the two FMs \cite{PhysRevApplied.14.044053,J_Cramer_spin_valve}, though the excitation mechanism was different.

\subsection{Spin current in the FM/AFM bilayer} 
\label{sec:spin_current}
In an insulator, lacking any  charge transport, purely magnonic spin currents can be triggered in a temperature gradient leading to a  continuous flow of magnons in the  sample, that can be detected on the colder side in a Hall-bar geometry \cite{Cramer2017}. The thermal gradient that we have used so far was a simplistic step function localized at a specific layer. In the following, we use a more realistic linear temperature gradient that allows the investigation of extended spin current profiles. This scenario replicates experimental profiles more realistically.

A spin current is defined quantum mechanically as \cite{Adachi_2013}
    \begin{equation} \label{eqn:quantum_SC}
    J_s = -\hbar\sum_k s_k^z v_k.
    \end{equation}
Here, $s_k^z$ is the $z$-component of the spin density $\mathbf{s}_k$ with wave vector $k$, $v_k$ is the corresponding group velocity, and $-\hbar$ is the angular momentum transferred by the spin current. In the classical limit, we have to focus on the spin-wave intensity and the corresponding observable which we will use in the following as the spin-wave spin current previously computed for a FM in \cite{Ritzmann2015Model-33804}. We extend this definition,
   \begin{equation}
    \label{eqn:FM_spin_current}
   \begin{aligned}
            I_{\textrm{FM}}(r_z,T) &=  J_{\textrm{FM}}a\langle S_x(r_z,T)S_y(r_{z}+a,T) \\
            &- S_x(r_{z}+a,T)S_y(r_z,T) \rangle, \\
        \end{aligned}
        \end{equation}
for an AFM as 
        \begin{equation}
        \begin{aligned}
        \label{eqn:AFM_spin_current}
            I_{\textrm{AFM}}(r_z,T) &= J_{\textrm{AFM}}a\langle S_{x,A}(r_z,T)S_{y,B}(r_z,T) \\
            &- S_{x,B}(r_z,T)S_{y,A}(r_z,T) \rangle.  \\
        \end{aligned}
        \end{equation}           
 This equation is based on the hard-axis spin components, i.e., the $x$ and $y$-components for a ground state spin configuration along the easy-axis $\hat{z}$, and $r_z$ represents the position of the magnetic unit cell. For the FM, the equal-time spatial correlation of the spin-wave components is used to define the spin current depicted in equation \ref{eqn:FM_spin_current}. The AFM needs special treatment because of its alternating spin structure and, thus, we compute the spin current using the convention in equation \ref{eqn:AFM_spin_current}, where the differing signs of the spin-wave components are taken care of by an inter-sublattice product of the spin-wave components. It is comprehensible, that in case of the AFM, the consecutive layers are from different sublattices, so the product is between the spin-components of the counter sublattices. Finally, we use again absorbing boundary conditions along the measured direction at the colder end of the system to ensure no reflection. 

\begin{figure}[t!]
\centering
\includegraphics[width=0.75\linewidth]{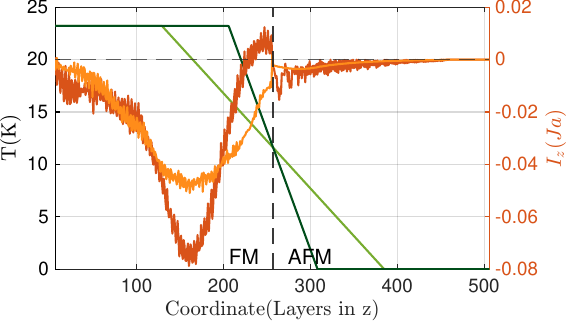}
   \caption{Spin-wave spin current profile (dark and light solid red lines) in an FM/AFM bilayer (dashed line designating the interface) with linear temperature  gradient shown as the dark and light green lines.}
   \label{fig:Tgrad_spin_current}
\end{figure}

Fig. \ref{fig:Tgrad_spin_current} shows the resulting spin current profile for the same model as in Figs. \ref{fig:spatial_FA} and \ref{fig:spatial_total} but for a linear temperature gradient.

The sign of the spin current is a consequence of the transport direction and the sign of the angular momenta being transferred. For the FM, the thermal gradient creates a spin current which carries negative angular momenta in the positive $z-$ direction. The asymmetry and the sign changes of $I_z$ in the FM near the interface are due to enhanced reflection of magnons at the interface. This is another signature of mode filtering as detected previously in the nonequilibrium magnon dispersions shown in Fig. \ref{fig:FM_AFM_2d} and \ref{fig:2D_AFM_FM}. In the AFM we detect a smaller spin current of the same sign that decreases exponentially.

Naively, one would expect a vanishing spin current in the AFM even inside the thermal gradient, since magnons of the two bands should compensate each other in the non-relativistic limit.
However, in this bilayer along with the usual magnons in the AFM there are additional ones injected from the FM which propagate into the AFM. This chirality selective injection of chiral RH-modes leads to the finite spin current in the AFM which vanishes away from the interface, indicating a proximity effect from the adjacent FM. 

\section{Conclusion}
We investigated equilibrium and non-equilibrium thermal excitation of magnons in a ferromagnet-antiferromagnet bilayer. In equilibrium, we calculated the magnon dispersion for this hybrid sample, where we find only right-handed modes in the FM but both, right and left-handed modes in the AFM. This serves as a reference for the investigation of non-equilibrium magnon excitations. 

To model a nonequilibrium thermal excitation in the form of a temperature gradient, we placed a  temperature step at the interface and investigated the propagation of magnons from the FM into the AFM and vice versa. The propagation is measured using two observables, the magnon accumulation and the magnonic spin current. We detected a finite magnon accumulation in both, the FM and the AFM peaking at the interface and decaying exponentially in the bulk. For the case of magnon injection from the FM into the AFM, a finite magnetization in the AFM is attributed to the mode selectivity which results from the unidirectional flow of RH-magnons from the FM into the AFM. This chiral mode selectivity due to the proximity to the interface generates a net magnetization in the AFM. The propagation lengths observed from our numerical results are a few tens of magnetic unit cells only which might seem insignificant for applications. However, the propagation length depends on the model  parameters, most importantly the damping parameter which was taken to be an optimal value to ensure smaller simulation dimension. In many materials, damping is much smaller than assumed here and thus propagation lengths will be significantly larger. To complement these findings, we computed the nonequilibrium magnon dispersion where we detect mode filtering at the interface owing to the directional and chiral preference of the FM on the AFM and vice versa. The proximity to the thermal gradient enhances this signal. These nonequilibrium magnon dispersions underpin the role of a chiral selectivity of the magnon modes transmitted in such heterostructures.

A more realistic linear temperature gradient is simulated to calculate the spin current traveling from the FM into the AFM. The sign of this current is preserved across the interface and its amplitude is clearly larger in the FM but is also detectable in the AFM. Hence, using this approach, one can induce a finite magnonic spin current into the AFM, without any relativistic effects, pointing potentially towards measurable spin currents in transport experiments.

\section{ACKNOWLEDGMENTS}
We acknowledge fruitful discussions with Levente R\'{o}zsa. This project was financially supported by the DFG through project No. 425217212 (SFB 1432, project B02). We are grateful for the computational resources provided by the Scientific Compute Cluster of the University of Konstanz (SCCKN).  

\section{DECLARATION}
There is no conflict of interest from any of the authors.
    
\section*{Credit authorship contribution statement}
The idea of the project was proposed by U.N. The simulations were performed by M.K. The manuscript was written by M.K. and revised by U.N. 

\section{DATA AVAILABILITY}
The data are not publicly available. The data are available from the authors upon reasonable request.

\sloppy  
\bibliography{Bibliography}
\fussy    
\end{document}